\documentstyle[pre,aps]{revtex}
\begin{document}
\draft
\title{Phase-ordering of conserved vectorial systems
with field-dependent mobility}

\author{Federico Corberi\thanks{corberi@na.infn.it}}
\address{
Dipartimento di Scienze Fisiche, Universit\`a di Napoli and
Istituto Nazionale di Fisica della Materia, Unit\`a di Napoli,
Mostra d'Oltremare, Pad. 19, 80125 Napoli, Italy}
\author{Claudio Castellano\thanks{claudio@ictp.trieste.it}}
\address{The Abdus Salam ICTP, Strada Costiera 11,
P.O. Box 586, 34100 Trieste, Italy}
\maketitle

\begin{abstract}

The dynamics of phase-separation in conserved systems with
an ${\cal O}(N)$ continuous symmetry is investigated in the presence 
of an order parameter dependent mobility $M\{\bbox \phi \}=1-a\bbox \phi ^2$.
The model is studied analytically in the framework 
of the large-$N$ approximation and by numerical simulations of the 
$N=2$, $N=3$ and $N=4$ cases in $d=2$, for both critical and off-critical 
quenches. We show the existence of a new universality class for $a=1$
characterized by a growth law of the typical length
$L(t)\sim t^{1/z}$ with dynamical exponent $z=6$ as opposed to the usual 
value $z=4$ which is recovered for $a<1$. 

\end{abstract}

\pacs{05.70.Fh, 64.60.Cn, 64.60.My, 64.75.+g}

\section{introduction}

The phase-separation kinetics of conserved systems quenched 
from a high temperature disordered state into the ordered region 
of the phase diagram is usually modeled by the Cahn-Hilliard \cite{cahn}
equation for the order parameter field 
$\bbox \phi (\bbox x,t)=\{\phi _\alpha (\bbox x,t)\}$, (with $\alpha =1,
\ldots, N$)
\begin{equation}
\frac{\partial \bbox \phi _\alpha (\bbox x,t)}{\partial t}=
\nabla \left \{ M(\bbox \phi )\nabla \left [ \frac{\delta 
F (\bbox \phi )}{\delta \bbox \phi _\alpha}\right ] \right \} 
\label{CH}
\end{equation}
In Eq.~(\ref{CH}) $F (\bbox \phi )$ represents an $O(N)$
symmetric free energy functional
with ground state in $\bbox \phi^2 /N = (1/N) \sum_{\alpha=1}^N
\phi^2_{\alpha} =1$ and $M(\bbox \phi )$ is the mobility.
Although in the original work of Cahn and Hilliard
a constant $M$ was proposed, it has been subsequently
argued \cite{kit} that, for scalar systems, a field-dependent mobility
\begin{equation}
M(\bbox \phi)\propto 1- a(T) \phi ^2
\label{mob}
\end{equation}
is more appropriate; $a(T) \to 1$ for temperature $T \to 0$, while
$a(T) \to 0$ for $T \to T_c$.
For vectorial systems it is natural to generalize Eq.~(\ref{mob}),
considering a mobility
\begin{equation}
M(\bbox \phi) \propto 1 - a(T) {\bbox \phi^2 \over N}
\label{mob3}
\end{equation}

Phase-ordering in presence of non-constant mobility has 
recently been studied with $N=1$\cite{lac,BP}, showing
a richer behavior with respect to the usual case with constant $M$.
From a microscopic point of view it is easy to understand the existence
of two different coarsening mechanisms in the separation of binary mixtures,
which is the typical realization of a scalar conserved phase-ordering
system.
The first is surface diffusion, namely the diffusion of molecules along
domain walls in order to minimize the interfacial energy, and is associated
with a growth law of the domain size $L(t)\sim t^{1/z}$, with $z=4$.
The second consists in the evaporation of molecules from large curvature
regions of domain interfaces and  their subsequent diffusion through
the other phase towards less curved interface portions.
It is often called also Lifshitz-Slyozov or evaporation-condensation
mechanism; the growth exponent is $z=3$\cite{LS}.
Since bulk diffusion is an activated process and surface diffusion is not,
an interesting phenomenology occurs when the temperature is changed.
For high $T$ (but still $T<T_c$) one observes only bulk diffusion ($z=3$),
which is faster than the other mechanism.
When the temperature is lowered, bulk diffusion is strongly suppressed
due to its activated nature; it is therefore possible to observe a 
preasymptotic regime dominated by surface diffusion ($z=4$).
This crossover is reproduced by the continuum equations with non constant
mobility\cite{lac,BP}, when the parameter $a$ is varied.
At the coarse-grained level of the continuum equations,
this again can be understood by observing that while for shallow quenches
($a \ll 1$) $M(\phi)$ remains finite in the whole system, the situation
is different when deep quenches are considered.
When $a=1$ the mobility vanishes within domains and suppresses bulk
transport; since $M(\phi)\simeq 1$ on domain boundaries, the motion of
interfaces is instead unaffected.

Despite the formal analogy of Eq.~(\ref{CH}) for scalar and
vectorial fields the underlying coarsening mechanisms
are in principle different, as a consequence of the different symmetry
of the ground state.
However for $N \leq d$ one has stable localized topological defects,
of which domain walls are the scalar counterpart.
With constant mobility defects play in vectorial systems exactly the same
role of surfaces for $N=1$: Phase-ordering proceeds for long times by
reduction in their typical radius of curvature (if they are extended)
or via mutual annihilation of defect-antidefect pairs (for point defects).
This analogy is however not complete: It is not clear what are in the
vectorial case the coarsening mechanisms corresponding to surface and bulk
diffusion; it is not even clear that two different mechanisms must exist
at all.
Furthermore, when $N>d$ no stable localized defects are present and the
coarsening process has evidently a different nature, even if the growth
exponent is the same.

In this paper we consider the phase-ordering kinetics of a vectorial field
with a non constant mobility.
We report results obtained numerically in $d=2$ for the case $N=2$,
as a paradigm 
of systems with topological defects, and the cases $N=3$ and $4$, where
stable defects are absent. We also study analytically the soluble large-$N$
model.
Interestingly, we find that the global picture is the same from $N=2$ up
to $N=\infty$, and perfectly analogous to the scalar case.
More precisely, while for $a\ll 1$ the power
growth law $L(t)\sim t^{1/z}$ is obeyed asymptotically with
$z=4$, with $a=1$ a new value of the dynamical 
exponent $z=6$ is observed asymptotically, corresponding to
a lower growth rate, in analogy with the case $N=1$.
For $a\lesssim 1$ the $z=6$ exponent is expected 
to be observed preasymptotically
before a crossover leads to $z=4$.
These results are found both for critical and off-critical quenches.

From these results it turns out that the similarity between ordering in
scalar and vectorial systems is very strong.
Also for $N>1$ the dynamic exponent $z$ is changed by the depth of the
quench in the presence of a non constant mobility.
This close analogy is not limited to fields supporting topological
defects but is instead valid for any number of components of the order
parameter.

The article is divided as follows:
In Sec.~2 the model is described;
in Sec.~3 the solution of the large-$N$ model is presented;
in Sec.~4 the results of numerical simulations
of two-dimensional systems with $N=2$, $N=3$ and $N=4$ are presented and
compared to the cases $N=1$ and $N=\infty$;
in Sec.~5 we summarize the results and draw our conclusions.

\section {The model}

We consider a system with a vectorial order parameter 
$\bbox \phi (\bbox x , t)$
initially prepared in a configuration sampled from an high temperature 
uncorrelated state with expectation values
$<\phi _\alpha (\bbox x ,0)>=N^{\frac{1}{2}}m_\alpha$ and 
$<\phi _\alpha (\bbox x,0)\phi _\beta (\bbox x',0)>=\Delta 
\delta _{\alpha \beta}\delta (\bbox x -\bbox x')$. For $t\geq 0$ the time 
evolution is governed by the noiseless Langevin equation~(\ref{CH}),
where $M(\bbox \phi)$ is given by Eq.~(\ref{mob3}),
and $F\{\bbox \phi\}$ is assumed in the
Ginzburg-Landau form:
\begin{equation}
F\{ \bbox\phi\}=\int d\bbox x \left[\frac {1}{2}|\nabla \bbox \phi|^2 -
\frac {1}{2} \bbox\phi ^2+\frac{1}{4N}( \bbox\phi ^2)^2\right]
\label {ham}
\end{equation}
With these positions the equation of motion for the order parameter
field reads
\begin{equation}
\frac {\partial \phi _\alpha(\bbox x,t)}{\partial t}=\nabla\left \{
\left [1-a \frac{\bbox \phi ^2(\bbox x,t)}{N}\right ]\nabla 
\left [-\nabla ^2 \phi _\alpha (\bbox x,t)
-\phi_\alpha (\bbox x,t)+\frac{1}{N}\bbox \phi ^2(\bbox x,t)
\phi _\alpha (\bbox x,t)\right ]\right \}
\label {eqmot}
\end{equation}
The central quantity in the study of the dynamical process described above is 
the structure factor, namely the Fourier transform of
the pair connected equal time correlation function, 
defined by:
\begin{equation}
C_\alpha(\bbox k,t)=<\phi_\alpha (\bbox k,t)
\phi _\alpha(-\bbox k ,t)>-Nm_\alpha ^2\delta(\bbox k)
\label {strucfac}
\end{equation}
In~(\ref{strucfac}) and hereafter $<...>$ means ensemble averages,
that is over the stochastic initial conditions, since
Eq.~(\ref{eqmot}) is deterministic.
If the $m_\alpha$ are equal (for critical quenches)
all the components of the structure factor are equivalent and the index
$\alpha $ will be dropped. 

\section {The large-$N$ model}

In this Section we present the analytical solution of the vectorial
model in the limit of an infinite number of components of
the order parameter field. 
Although it has been shown \cite{ccz97} that the nature of the 
dynamical process is different when $N$ is strictly infinite, the
signature of this being the multiscaling symmetry
obeyed by the structure factor, the global picture provided by
this model is often qualitatively adequate and predicts 
the correct dynamical exponent $z$ for $N>1$.
Here it will be shown that the non trivial behavior of
systems with non constant mobility, which is due to the
interplay between the different coarsening mechanisms, 
is fully reproduced by the large-$N$
model.

Let us consider a process where symmetry breaking can be induced along
one direction by virtue of an asymmetric initial condition
\begin{equation}
m_1\neq 0 \hspace {1 cm}\,\,\, m_\beta=0 \,\,\,\,\,\,\,\,\,\{\beta=2,...,N\}
\end{equation}

In the large-$N$ limit
the evolution
equations for the longitudinal and transverse components of the 
structure factor (see Appendix) read

\begin{equation}
\frac{\partial C_1 (\bbox k,t)}{\partial t}=
2\{1-a[S_\bot(t)+m_1^2]\}k^2\left [-k^2+1-S_\bot(t)-3m_1^2)\right ]
C_1(\bbox k,t)
\label{N=oolong}
\end{equation}
\begin{equation}
\frac{\partial C_\bot (\bbox k,t)}{\partial t}=
2\{1-a[S_\bot(t)+m_1^2]\}k^2\left [-k^2+1-S_\bot(t)-m_1^2)\right ]
C_\bot(\bbox k,t)
\label{N=oo}
\end{equation}
where $C_1$ refers to the correlations along the symmetry breaking
direction, $\alpha=1$, and $C_\bot$ is the structure factor along
one of the equivalent transverse directions.
$S_\bot(t)$ can be computed self-consistently through
\begin{equation}
S_\bot(t)=\int _{|{\bf k}|<q}\frac{d{\bf k}}{(2\pi)^d}C_\bot({\bf k},t)
\label{eqS}
\end{equation}
where $q$ is a phenomenological ultraviolet momentum cutoff.

Eq.~(\ref{N=oo}) together with the self-consistency relation (\ref{eqS})
governs the dynamics in the large-$N$ model.
These equations apply to off-critical and critical quenches ($m_1=0$)
as well; in the latter case only one equation is required since 
Eqs.~(\ref{N=oolong}) and (\ref{N=oo}) coincide. 
Notice that Eq.~(\ref{N=oo}) for $C_\bot$ does not contain $C_1$ and
therefore,
with an high temperature disordered initial condition
$C_\bot({\bf k},0)=\Delta$ can be formally integrated
yielding
\begin{equation}
C_\bot(\bbox k,t)=\Delta e^{-k^4\Lambda ^4(t)+k^2{\cal L}^2(t)}
\label{eqint}
\end{equation}
where the lengths
\begin{equation}
\Lambda (t)=\left \{ 2\int _0 ^t \{1-a[S_\bot(\tau)+m_1^2]\}d\tau  \right \} 
^{1/4}
\label{lambda}
\end{equation}
\begin{equation}
{\cal L} (t)=\left \{ 2\int _0 ^t \{1-S_\bot(\tau)-m_1^2\} 
\{1-a[S_\bot(\tau)+m_1^2]\} d\tau  \right \} ^{1/2}
\label{call}
\end{equation}
have been introduced.
For long times the structure factor is sharply peaked around
\begin{equation}
k_m=\frac {{\cal L}(t)}{\sqrt {2}\Lambda ^2 (t)}
\end{equation}
allowing a saddle point evaluation of the integral over momenta in
Eq.~(\ref{eqS}) which yields
\begin{equation}
S(t)\sim \Delta {\cal L}^{-d}(t)
e^{\frac{{\cal L}^4(t)}{4\Lambda ^4 (t)}}
\label{saddle}
\end{equation}
For long times convergence towards thermodynamic equilibrium requires
the order parameter to approach the minimum of the local part
of the free energy (\ref{ham}) which is for $\bbox \phi ^2=N$.
Hence, letting $S_\bot(t)\simeq 1-m_1^2$ asymptotically in Eq.~(\ref{saddle})
one has 
\begin{equation}
\Lambda (t)\simeq \frac{{\cal L}(t)}{[4d\ln {\cal L}(t)]^{\frac{1}{4}}}
\label{saddle2}
\end{equation}
Two different cases must then be considered, namely $a<1$ or $a=1$.
For $a<1$ one immediately obtains from Eq.~(\ref{lambda})
$\Lambda (t)\sim t^{1/4}$ which yields 
${\cal L}(t)\sim (t\ln t)^{1/4}$ due to Eq.~(\ref{saddle}).
Hence the physical length $L(t)\sim k^{-1}_m (t)$ 
associated to the peak of the structure
factor grows as $L(t)\sim [t/\ln (t)]^{1/4}$.
With $a<1$, therefore, the asymptotic behavior is the same as
for a large-$N$ system with constant mobility \cite{crz},
as expected. Notice the logarithmic correction with
respect to the power law growth obeyed by systems
with finite $N$ which is due to the multiscaling symmetry of
$C(\bbox k,t)$; the dynamical exponent $z=4$ is known to be correct
for all physical vectorial systems. 

Let us consider now the case $a=1$.
In this case, by matching Eqs.~(\ref{lambda}), (\ref{call}) and
(\ref{saddle2}) one finds a different solution characterized
by $\Lambda (t)\sim t^{1/6}$ and ${\cal L}(t)\sim t^{1/6}
(\ln t)^{1/4}$, which yields $L(t)\sim t^{1/6}/(\ln t)^{1/4}$.
With $a=1$, therefore the vanishing of the mobility in equilibrated
regions slows down the dynamics and changes $z$ from 4 to 6, similarly to
the scalar case where one goes from the Lifshitz-Slyozov
evaporation condensation mechanism, associated to $z=3$, to $z=4$. 
It must be noticed however that,
apart from this analogy, the physics of the coarsening process
is very different due to the absence of stable topological defects for
$N>d$ and there is no clear indication on the nature of the transport
mechanism associated with $z=6$.

\section{Systems with finite $N$}

In this section we present the results of the numerical solution
of Eq.~(\ref{eqmot}).
Our aim is to present a rather complete description of the
effect of a non-constant mobility in systems with different
number of components and to compare the results with what is
known for $N=1$ and with the above discussed large-$N$ model.
In order to fulfill this program we have chosen $d=2$ 
and $N=2$ or $N=3$ and $4$ so that we can analyze systems with and
without stable topological defects.

The numerical solution of Eq.~(\ref{eqmot}) is obtained by simple
iteration  of the discretized equation on a 512x512 mesh, except
for the off-critical $N=2$, $a=1$ case which is 
computed on a 256x256 lattice. Each quench is averaged over three
different realizations of the initial conditions.
The numerical solution of  Eq.~(\ref{eqmot}) is particularly 
delicate for $a\lesssim 1$, because in this case a small error
in the calculation of $\bbox \phi$ in the bulk may give rise to
a negative $M(\bbox \phi)$, thus producing a spurious instability.
We have avoided this problem by letting $M(\bbox \phi)=|1-a\bbox \phi ^2|$,
and tested that this does not produce any significative difference in the
results.
 
The characteristic growing length $L(t)$ is obtained from
the structure factor as $L(t)=k_1^{-1}(t)$, where
\begin{equation}
k_1(t)={\int dk k C(k,t)\over
\int dk C(k,t) }
\end{equation}
is the first moment of $C(k,t)$.

\subsection {$N=3$ and $N=4$}

In Fig.~(\ref{1}) the behavior of $L(t)$ is compared in the
two cases $a=0$ and $a=1$, for critical quenches with $N=3$ and $N=4$. 
After the linear instability is over (which
corresponds to times up to $\ln(t)\simeq 4$),
the system enters the asymptotic regime which is characterized by
a power law growth of $L(t)$. For $a=0$ one has $z=4$ with
good accuracy almost immediately, while for $a=1$ a  
convergence towards $z=6$ is observed. 
Best fit estimates for times larger than $t=3000$ yield $z=3.92$ and 
$z=3.89$ for $a=0$ and $N=3$ and $N=4$ respectively;
for $a=1$, instead, one has $z=5.68$ and $z=5.81$ 
in the corresponding cases.
We notice, by the way, the remarkable superposition of the
curves with $N=3$ and $N=4$.

It is interesting to observe the evolution of the structure factor,
which is shown in Figs.~(\ref{2}) and (\ref{3}), 
for $N=3$, at different times for $a=0$
and $a=1$, respectively.
Here one observes, immediately after the linear instability corresponding
to the exponential growth of a peak at constant wave vector, the 
formation of a rather well
developed power law decay of $C(\bbox k,t)$ for large $k$; 
this tail is initially formed at very large values of $k$, subsequently it
moves towards lower wave vectors and then
gradually disappears as time increases, being replaced by
a faster (exponential) decay of the structure factor.
As Fig.~(\ref{2}) shows the exponent of this decay is very close to $5$
in accordance with the existence of a generalized Porod's tail, i.e.
$C(\bbox k,t)\sim k^{-(N+d)}$ \cite{bpt}.
It is well known that the Porod's tail is associated with
the presence of localized topological defects.
The presence of such a tail in Figs.~(\ref{2}) and (\ref{3}),
suggests therefore that defects are formed at early time and eventually
decay so that the power law disappears asymptotically.
This phenomenon is displayed, with similar features, both for $a=0$ and
for $a=1$, and will be studied elsewhere\cite{ccz98}.

In Figs.~(\ref{4})  and (\ref{5}) the data collapse for
$C(\bbox k,t)$ are presented.
Here 5 curves corresponding to about $1.5$ decades in time are superimposed
by plotting $L(t)^{-d}C(\bbox k,t)$
against $x=kL(t)$. We observe that for $a=0$ the data collapse is very good
for $x\lesssim 2$, whereas it becomes progressively less accurate for
increasing
values of $x$. This effect is probably due to the remnant of the
power law tail that, for the times covered by the numerical solution, has not
completely disappeared, as can be seen from Fig.~(\ref{2}).
In any case the data are consistent with 
the existence of the scaling property both for $a=0$ and
$a=1$. In Fig.~(\ref{6}) the scaling functions of the cases $a=0$ and $a=1$
are compared. 
The scaling function is obtained here as $L(t_M)^{-d}C(\bbox k,t_M)$,
$t_M$ being the longest time of our computation.
It has been shown in \cite{BP}
that no significative differences are observed in the scalar case
between the two scaling functions, a fact that indicates 
the independence of the morphology of the growing phases
with respect to a change of the coarsening mechanism
from bulk to surface diffusion.
Fig.~(\ref{6}) shows that the same property is shared by our vectorial
model.
Notice also that $C(\bbox k,t)\sim k^4$ for small $k$~\cite{yeung}.

\subsection{$N=2$}

We present in the following the results of the numerical
simulation of the case $N=2$.
In Fig.~(\ref{7}) the behavior of $L(t)$ is plotted versus $t$ in the
two cases $a=0$ and $a=1$.
From this figure we conclude that
a power law growth with $z=4$ or $z=6$ for $a=0$ and
$a=1$ respectively, fits very well the data.
Linear regression analysis for $t>3000$ yields $z=4.20$ and $z=6.26$ 
for $a=0$ and $a=1$, respectively.

The structure factors at different times for $a=0$
and $a=1$ are shown in Figs.~(\ref{9}) and (\ref{10}), respectively.
Here one observes again the formation of 
the Porod's tail but, differently from the cases $N=3$ and $N=4$, 
this pattern is maintained asymptotically, a fact that reflects the
stability of the topological defects in the late stage of the dynamics. 
In Fig.~(\ref{11})  and (\ref{12}) the scaling plots for
$C(\bbox k,t)$ are presented.
No compelling evidence can be obtained by the analysis of this figures
about the possible existence of a scaling breakdown,
which has been suggested with constant mobility for $d=2$
and $N=2$\cite{Bray94}; nevertheless
one observes a worse data collapse with respect to the case 
$N=3$ and $4$ for large $x$, both for $a=0$ and $a=1$.
For $x\lesssim 2$, on the other hand, the collapse is rather good. 
In Fig.~(\ref{13}) a comparison is presented between the 
quantity $L(t_M)^{-d}C(\bbox k,t_M)$ (which, if scaling exists,
corresponds to the scaling function) in the cases $a=0$ and $a=1$.
Here one observes a superposition of the two curves for $x\lesssim 2$.
For $x>2$, however, they become more and more distant with $x$.
The situation is different from the cases $N=1$ and $N>2$ where
the scaling functions practically coincide in the whole range of $x$ values.
From the observation of Figs.~(\ref{11}),(\ref{12}) and (\ref{13}) 
we argue that,
for $x\lesssim 2$ the $N=2$ model behaves very similarly to the
cases $N=1$ and $N>2$, since we find a superposition of curves at different
times in the scaling plots (\ref{11}), (\ref{12}), 
and very similar  "scaling function"
for $a=0$ and $a=1$. For $x>2$, however, one observes a much worse 
superposition in both cases. It is not presently clear if 
the origin of this difference is due to preasymptotic effects in the
simulations or to scaling violations.

In Fig.~(\ref{14}) the effect of a finite $m_\alpha$ is considered and 
the growth of the characteristic length $L(t)$ is 
shown vs. $t$ for $a=1$ both for critical and off-critical quenches.
The asymmetric quench is realized by taking 
$m_\alpha =0.65/\sqrt{2}, \forall \alpha$.
The solution for the off-critical case has been obtained on a 256x256
lattice in order to speed up the computation since, as the figure shows, 
much longer times than in the critical case 
must be reached in order to observe the asymptotic behavior.
For sufficiently long times one observes $z=6$ in both cases, suggesting
that the same coarsening mechanism is at work both for critical and
off-critical quenches.

\section {Conclusions}

In this paper we have considered the effect of an order parameter dependent
mobility on the phase-ordering process of a vectorial system quenched
in the ordered region of the phase diagram.
This problem has been studied both numerically, by considering the cases $N=2$,
$N=3$ and $N=4$, and analytically in the large-$N$ limit, for critical and 
off-critical quenches. We have shown that for $a=1$ the dynamics is slowed
down with respect to the case $a=0$, due to the vanishing of the mobility 
in pure phases. A power law growth of the typical length $L(t)\sim t^{1/z}$
is still obeyed in this case but with $z=6$.
This behavior is exhibited in vectorial systems for all the values of $N$
considered and does not depend on the symmetry of the quench, nor on the
presence of topological defects.
We conclude that this case falls within a different universality class
with respect to $a=0$.
For intermediate values of $a$ the system behaves as if $a=0$ asymptotically 
but a preasymptotic growth with $z=6$ is expected for $a$ sufficiently 
close to 1.

These features are very reminiscent of those of scalar systems,
where the dynamical exponent changes from $z=3$ to $z=4$ when $a=1$.
It is however important to point out that the scalar models studied
in Ref.\cite{lac} and \cite{BP} are not perfectly equivalent.
The difference is in the form of the free energy functional which is
of the Ginzburg-Landau type (quartic polynomial) in Ref.\cite{lac}
and of logarithmic type in \cite{BP}.
While the polynomial can be seen as the expansion of the logarithmic
form for $T$ not too far from $T_c$, for lower temperatures only the
logarithmic form provides a correct coarse-grained description of the
microscopic dynamics.
This is confirmed by what happens for off-critical quenches to $T=0$.
When the quench is sufficiently asymmetric so that the minority phase
forms a non percolating pattern of isolated droplets, surface diffusion
alone cannot drive the phase-separation to completion.
The system therefore remains pinned in a configuration out of equilibrium.
This pinning is correctly reproduced by the continuum equation
with the logarithmic free energy\cite{Castellano98}; no such pinning
is found with the quartic polynomial\cite{BE}.
In this last case the deep off-critical condition has only the effect
of changing bulk-diffusion to a slower mechanism ($z=4$) which has been
referred to as bulk sub-diffusion.
The dynamical exponent for bulk sub-diffusion turns out to be the same 
as that for surface diffusion: We do not know whether this equality
is accidental or has a deeper meaning.

In our study we have considered the generalization to a vectorial order
parameter of the model with Ginzburg-Landau free energy.
Also in this case we find two universality classes for $a=1$ and $a<1$.
This is perfectly analogous to what happens in the scalar case.
Moreover the scaling functions for $a=0$ and $a=1$ are identical.
This leads us to conjecture that a mechanism of the same nature drives
phase-ordering for all values of $a$. The only special feature of the
$a=1$ case is that the mechanism is slowed down, exactly as for $N=1$ one
goes from bulk diffusion to bulk sub-diffusion, but the nature of the
mechanism is still of bulk type.

It would be interesting to check whether a sort of generalized surface
diffusion exists for $N>1$. This could be detected by solving the
equation of motion for $a=1$ with a logarithmic free energy for a
vectorial order parameter.
An indication that this could be the case is provided by the analytical 
solution of the large-$N$ model with a logarithmic form of the free
energy\cite{Castellano98b}. 

\section*{Acknowledgments}

F. C. is very grateful to M. Cirillo for hospitality.

\section*{Appendix}

Defining the fluctuation field $\psi(\bbox x,t)$ as
\begin{equation}
\phi_1(\bbox x,t)=N^{\frac{1}{2}}m_1+\psi (\bbox x,t)
\end{equation}
and inserting into Eq.~(\ref{eqmot}) one obtains the pair of equations
\begin{eqnarray} 
\frac{\partial[N^{\frac{1}{2}}m_1+\psi(\bbox x,t)]}{\partial t} 
     &=&
             \nabla \left \{ \{1-a[\frac{1}{N}\sum_{\beta=2}^N\phi^2_\beta 
             (\bbox x,t)+m_1^2+\psi^2(\bbox x,t)+
             2N^{\frac{-1}{2}}m_1\psi(\bbox x,t)]\}  \right . \nonumber \\ 
     & &     \nabla \{
             -\nabla ^2 \psi (\bbox x,t)-N^{\frac{1}{2}}m_1-\psi(\bbox x,t)+
             \frac{1}{N}[N^{\frac{1}{2}}m_1+\psi (\bbox x,t)]
             \sum_{\beta=2}^N\phi^2_\beta (\bbox x,t)+ \nonumber \\
     & &     \left . N^{\frac{1}{2}}m_1^3+3m_1^2\psi(\bbox x,t)+ 
             \frac{3}{N^{\frac{1}{2}}}m_1\psi^2(\bbox x,t)
             +\frac{1}{N}\psi^3(\bbox x,t)\} \right \} 
\end{eqnarray}
\begin{eqnarray}
\frac{\partial[\phi_\beta(\bbox x,t)]}{\partial t}
     &=&
          \nabla \left \{ \{1-a[\frac{1}{N}\sum_{\beta=2}^N
          \phi^2_\beta (\bbox x,t)+m_1^2+\psi^2(\bbox x,t)+
          2N^{\frac{-1}{2}}m_1\psi(\bbox x,t)]\} \nabla \{ -\nabla ^2 
          \phi_\beta (\bbox x,t) \right . \nonumber \\
     & &  \left . - \phi_\beta (\bbox x,t)+ 
          \frac{1}{N}\sum_{\alpha=2}^N\phi^2_\alpha (\bbox x,t)
          \phi_\beta (\bbox x,t)
          +[m_1^2+\frac{2}{N^{\frac{1}{2}}}m_1\psi(\bbox x,t)
          +\frac{1}{N}\psi^2(\bbox x,t)]
          \phi_\beta(\bbox x,t) \right \}
\end{eqnarray}
with $\beta \neq 1$.

In the large-$N$ limit, summing over vector components averages
the system over an ensemble of configurations and hence
\begin{equation}
\lim _{N\to\infty}\frac{1}{N}|\bbox \phi (\bbox x,t)|^2=
\lim _{N\to\infty}\frac{1}{N}\sum _{\beta =2}^N
\phi ^2_\beta (\bbox x,t)=<\phi ^2_\beta (\bbox x,t)>\equiv
S_\bot(t)
\end{equation}
where translational invariance has been assumed and $S_\bot(t)$
does not depend on $\beta$ due to the internal symmetry.
To leading order in $N$ one obtains:

\begin{equation}
\frac{\partial \psi (\bbox x,t)}{\partial t}=
\{1-a[S_\bot (t)+m_1^2]\}\nabla ^2\left \{ [-\nabla^2-1+S_\bot (t)+3m_1^2]
\psi (\bbox x,t)\right \}
\label{fieldlong}
\end{equation}

\begin{equation}
\frac{\partial \phi _\beta (\bbox x,t)}{\partial t}=
\{1-a[S_\bot (t)+m_1^2]\}k^2\left \{ [-\nabla^2-1+S_\bot (t)+m_1^2]
\phi _\beta (\bbox x,t)\right \}
\label{fieldtrasv}
\end{equation}

Introducing the longitudinal and transverse part of the structure factor, 
namely $C_1(k,t)=<\psi(\bbox k,t)\psi(-\bbox k,t)>$ and 
$C_\bot(k,t)=<\phi _\beta(\bbox k,t)\phi _\beta(-\bbox k,t)>$, which is 
independent on $\beta$ due to internal symmetry 
and Fourier transforming one obtains the following pair of equations

\begin{equation}
\frac{\partial C_1 (\bbox k,t)}{\partial t}=
2\{1-a[S_\bot(t)+m_1^2]\}k^2 [-k^2+1-S_\bot(t)-3m_1^2]
C_1(\bbox k,t)
\label{appN=oolong}
\end{equation}

\begin{equation}
\frac{\partial C_\bot (\bbox k,t)}{\partial t}=
2\{1-a[S_\bot(t)+m_1^2]\}k^2 [-k^2+1-S_\bot(t)-m_1^2]
C_\bot(\bbox k,t)
\label{appN=oo}
\end{equation}

\begin{figure}
\caption{
The growth law of the characteristic length $L(t)$ is shown for
systems with $N=3$ and $N=4$. Straight lines represent power 
laws $t^{1/z}$ with $z=4$ and $z=6$.
}
\label{1}
\end{figure}

\begin{figure}
\caption{
The evolution of $C(\bbox k,t)$ is shown at different times for
$a=0$ and $N=3$.
The straight line represents a power law decay $k^{-5}$. 
}
\label{2}
\end{figure}

\begin{figure}
\caption{
The evolution of $C(\bbox k,t)$ is shown at different times for
$a=1$ and $N=3$.
The straight line represents a power law decay $k^{-5}$. 
}
\label{3}
\end{figure}

\begin{figure}
\caption{
Data collapse (scaling plot) for $a=0$ and $N=3$. 
Here $L(t)^{-d}C(\bbox k,t)$ is plotted against $x=kL(t)$ for 
different times.
}
\label{4}
\end{figure}

\begin{figure}
\caption{
Data collapse (scaling plot) for $a=1$ and $N=3$. 
Here $L(t)^{-d}C(\bbox k,t)$ is plotted against $x=kL(t)$ for 
different times.
}
\label{5}
\end{figure}

\begin{figure}
\caption{
Comparison between the scaling functions for $N=3$
and $a=0$ or $a=1$. The straight line represents the power law
$k^4$.
}
\label{6}
\end{figure}

\begin{figure}
\caption{
The growth law of the characteristic length $L(t)$ is plotted 
against time for
a system with $N=2$. Straight lines represent power 
laws $t^{1/z}$ with $z=4$ and $z=6$.
}
\label{7}
\end{figure}

\begin{figure}
\caption{
The evolution of $C(\bbox k,t)$ is shown at different times for
$a=0$ and $N=2$.
The straight line represents a power law decay $k^{-4}$. 
}
\label{9}
\end{figure}

\begin{figure}
\caption{
The evolution of $C(\bbox k,t)$ is shown at different times for
$a=1$ and $N=2$.
The straight line represents a power law decay $k^{-4}$. 
}
\label{10}
\end{figure}

\begin{figure}
\caption{
Data collapse (scaling plot) for $a=0$ and $N=2$. 
Here $L(t)^{-d}C(\bbox k,t)$ is plotted against $x=kL(t)$ for 
different times.
}
\label{11}
\end{figure}

\begin{figure}
\caption{
Data collapse (scaling plot) for $a=1$ and $N=2$. 
Here $L(t)^{-d}C(\bbox k,t)$ is plotted against $x=kL(t)$ for 
different times.
}
\label{12}
\end{figure}

\begin{figure}
\caption{
Comparison between the "scaling functions" for $N=2$
and $a=0$ or $a=1$. The straight line represents the power law
$k^4$.
}
\label{13}
\end{figure}

\begin{figure}
\caption{
The growth law of the characteristic length $L(t)$ is plotted vs. 
$t$ for a system with $N=2$, for $a=1$, for critical and
off critical quenches. The straight line represents the power 
law $t^{1/z}$ with $z=6$.
}
\label{14}
\end{figure}


\begin{references}

\bibitem{cahn}
J.W. Cahn and J.E. Hilliard,
J. Chem. Phys. {\bf 28}, 258 (1958);
J.W.Cahn, Acta Metal. {\bf 9}, 795 (1961);
Trans. Met. Soc. AIME {\bf 242}, 166 (1968);
H.E. Cook,
Acta Metall. {\bf 18}, 297 (1970).

\bibitem{kit}
K. Kitahara and M. Imada, Suppl. Prog. Theor. Phys. {\bf 64}, 65 (1978);
J.S.Langer, M.Bar-on, H.D.Miller, Phys. Rev. A {\bf 11}, 1417 (1975).

\bibitem{lac}
A.M.Lacasta, A.Hern\'andez-Machado, J.M.Sancho and R.Toral,
Phys. Rev. {\bf E 45}, 5276 (1992).

\bibitem{BP} 
S.Puri, A.J.Bray, J.L.Lebowitz, Phys. Rev. {\bf E} 56, 758 (1997).

\bibitem{LS}
I.M. Lifshitz and V.V. Slyozov, J.Phys. Chem.Solids {\bf 19}, 35 (1961);
C.Wagner, Z.Electrochem. {\bf 65}, 581 (1961).

\bibitem{ccz97}
C. Castellano, F. Corberi and M. Zannetti,
Phys. Rev. {\bf E} 56, 4973, (1997).

\bibitem {crz}
A.Coniglio, P.Ruggiero and M. Zannetti, Phys. Rev. {\bf E} 50,
1046, (1994).

\bibitem{bpt}
A.J.Bray and S.Puri, Phys. Rev. Lett. 67, 2670, (1991);
H.Toyoki, Phys. Rev. {\bf B} 45, 1965, (1992).

\bibitem{ccz98}
C. Castellano, F. Corberi and M. Zannetti,
in preparation.

\bibitem{yeung}
C.Yeung, Phys. Rev. Lett. {\bf 61}, 1135, (1988).

\bibitem{Bray94}
A. J. Bray,
Adv. Phys. {\bf 43}, 357 (1994).

\bibitem{Castellano98}
C. Castellano and F. Corberi,
(submitted to J. Chem. Phys.).

\bibitem{BE} A.J. Bray and C.L. Emmott, Phys. Rev. {\bf B 52}, 685
(1995).

\bibitem{Castellano98b}
C. Castellano and F. Corberi,
Phys. Rev. {\bf E 57}, 672 (1998).

\end{references}
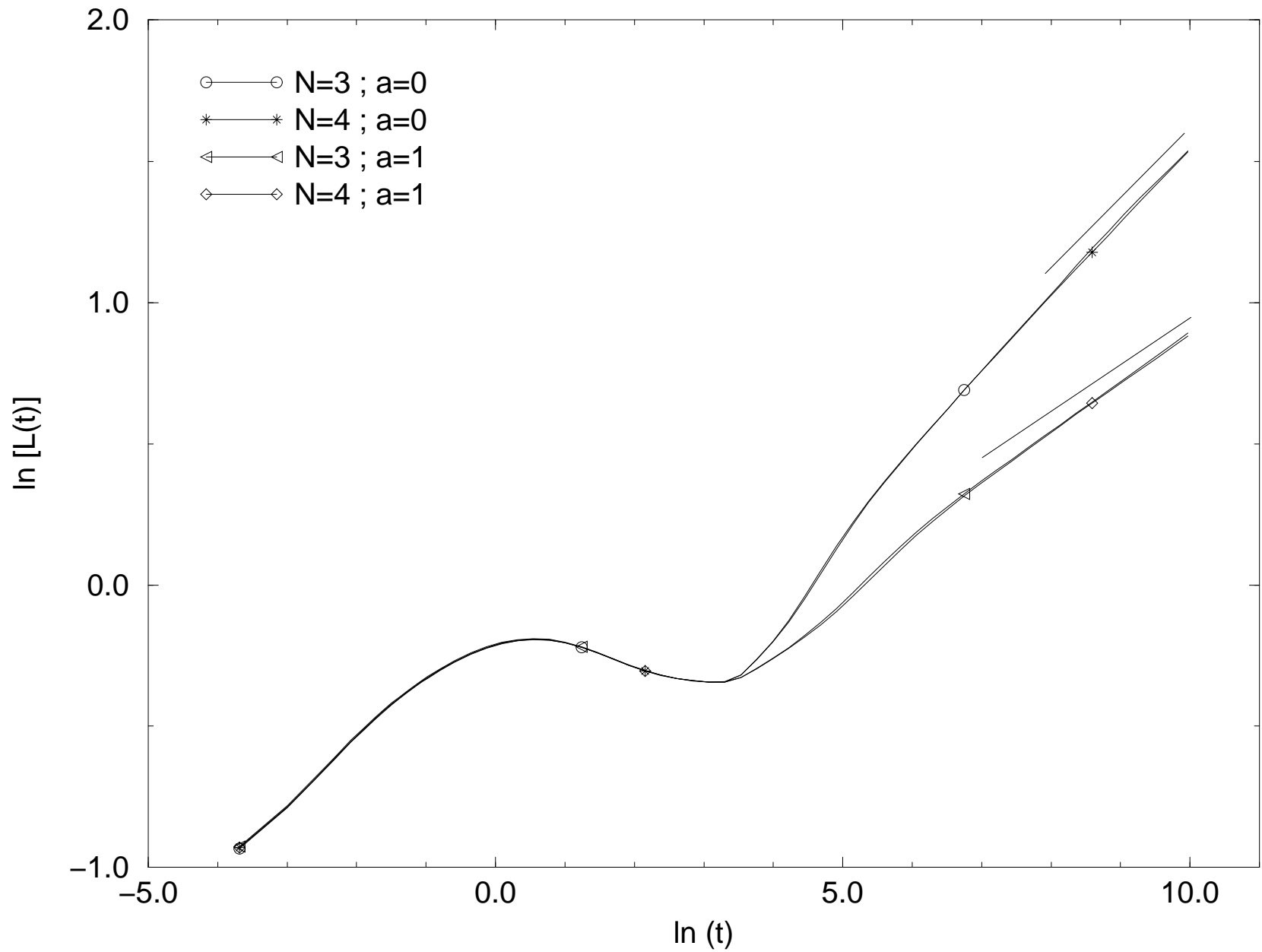
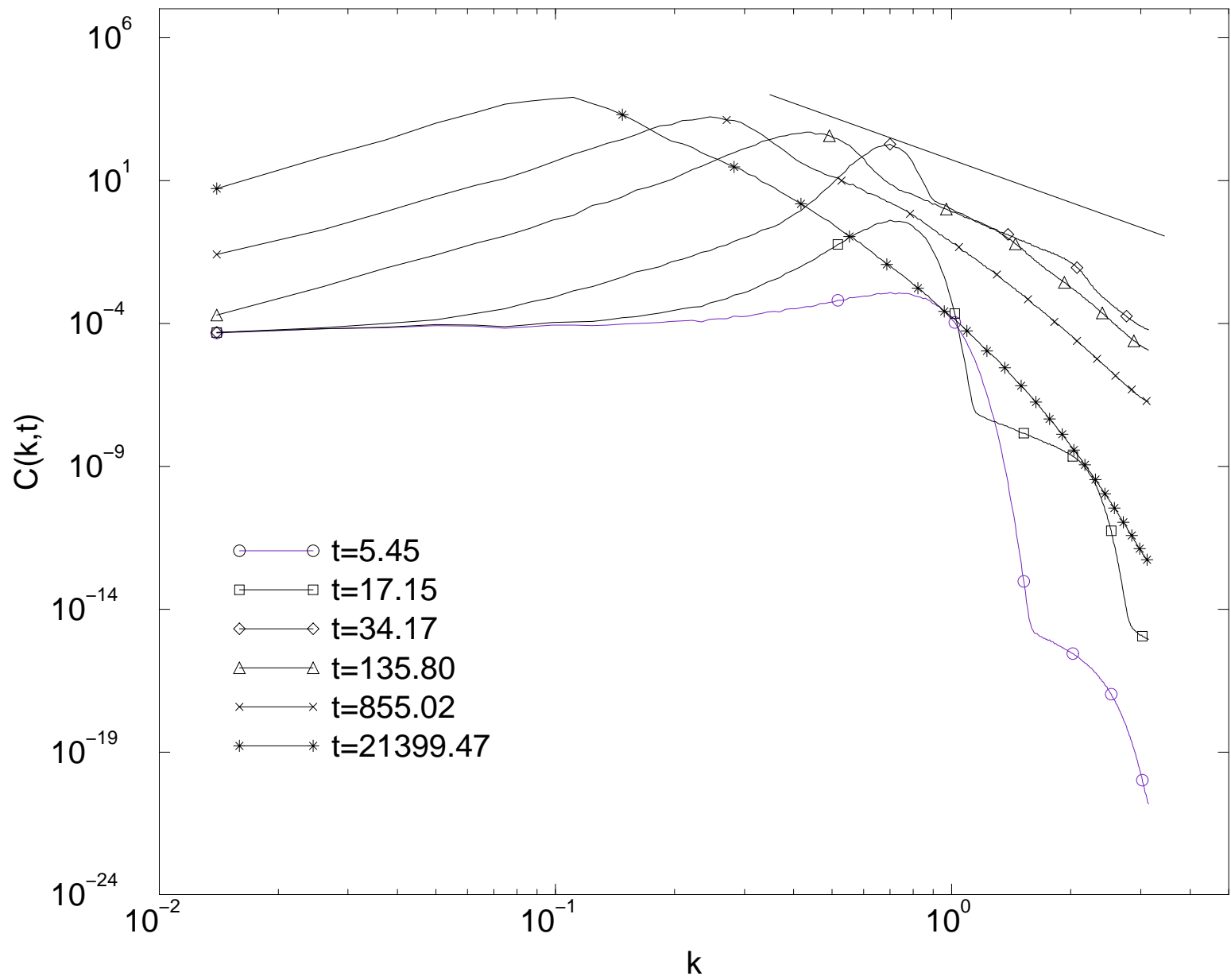
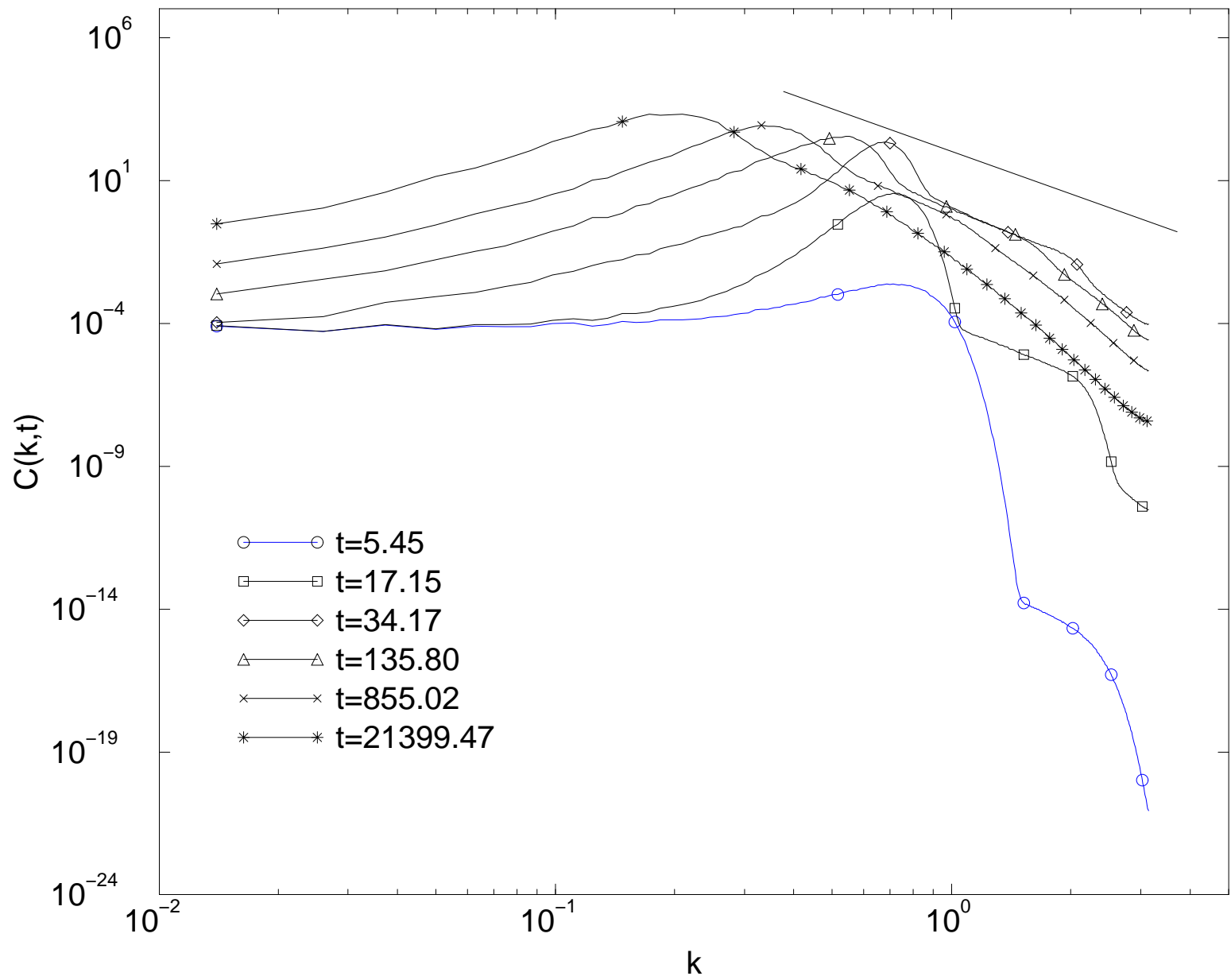
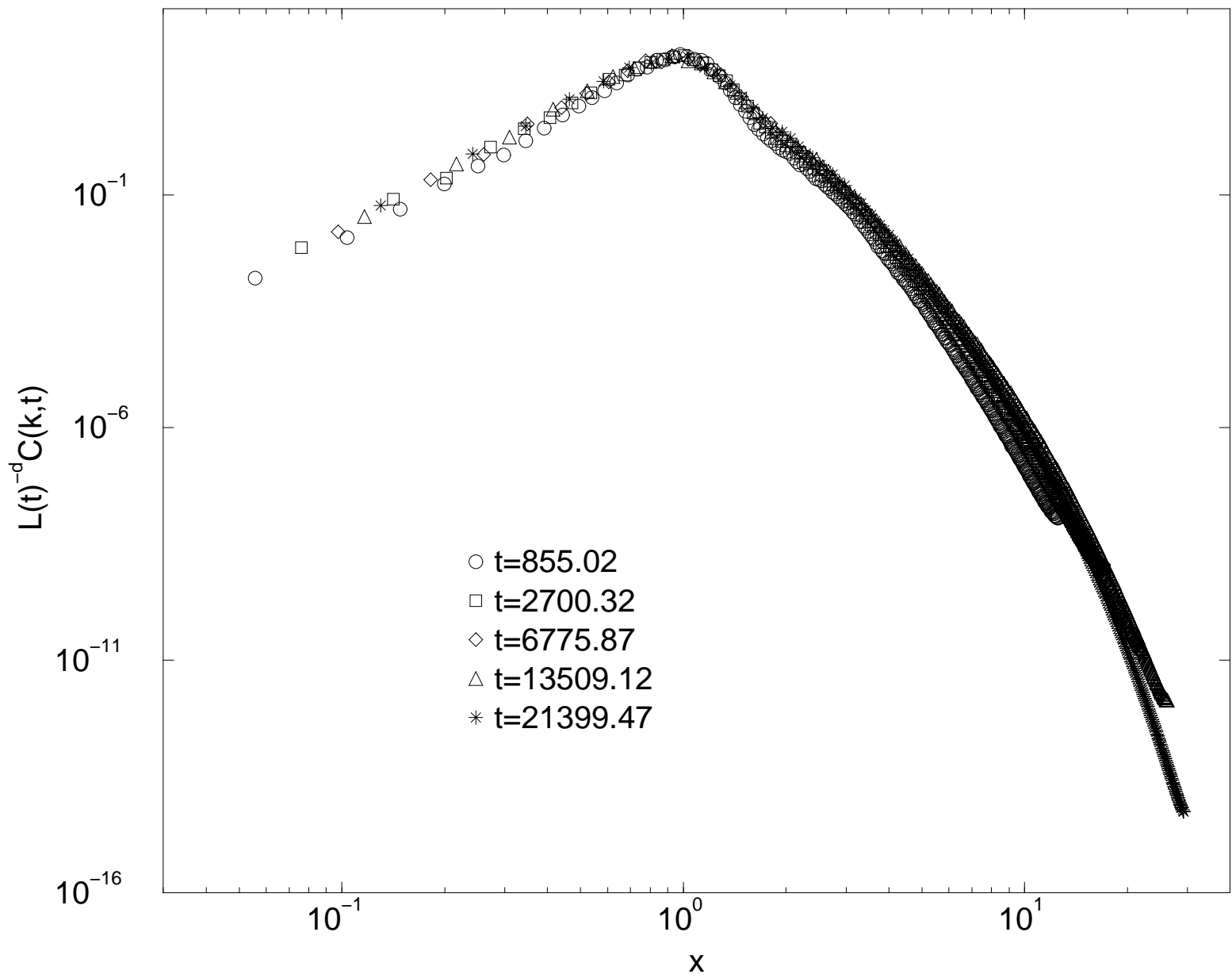
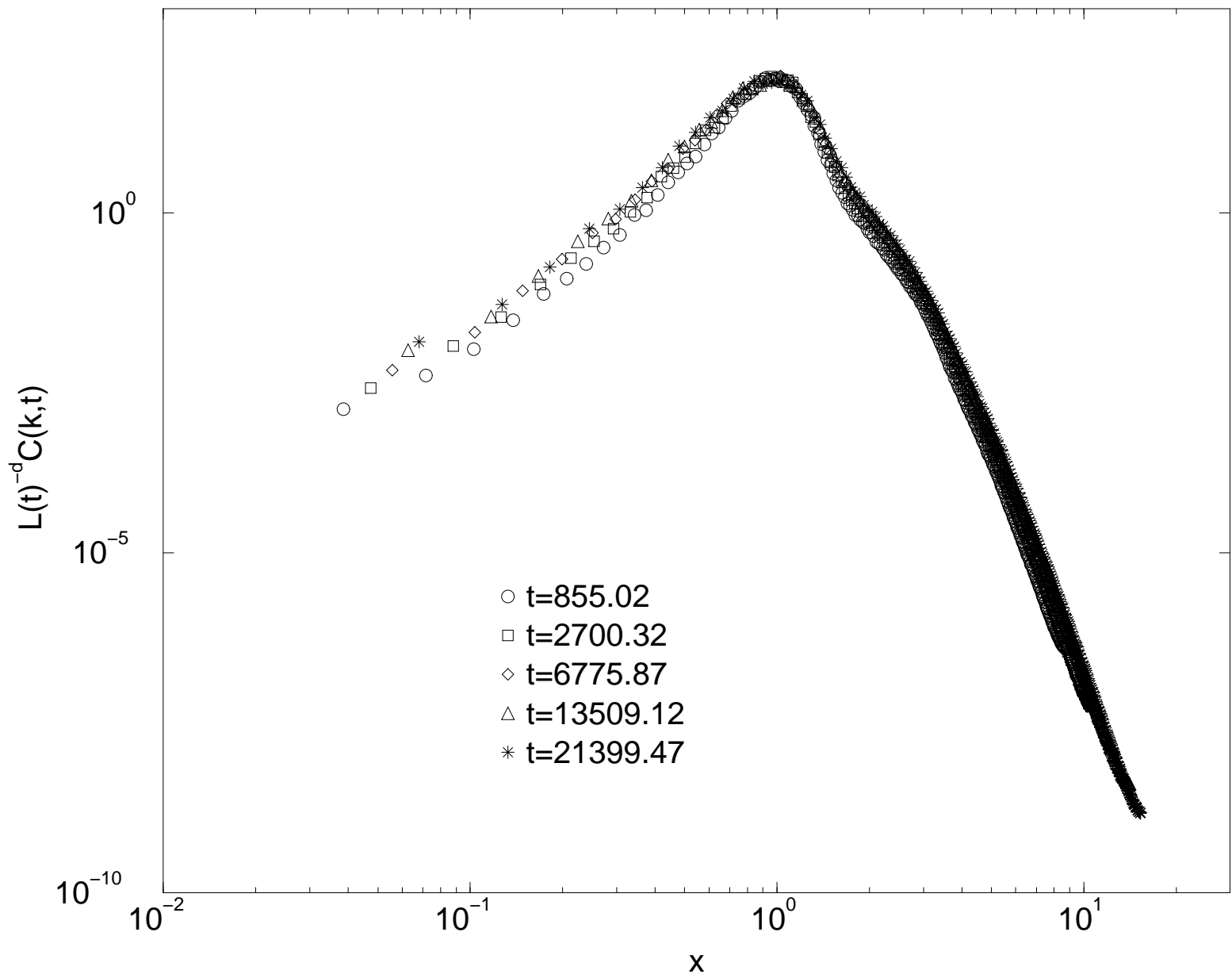
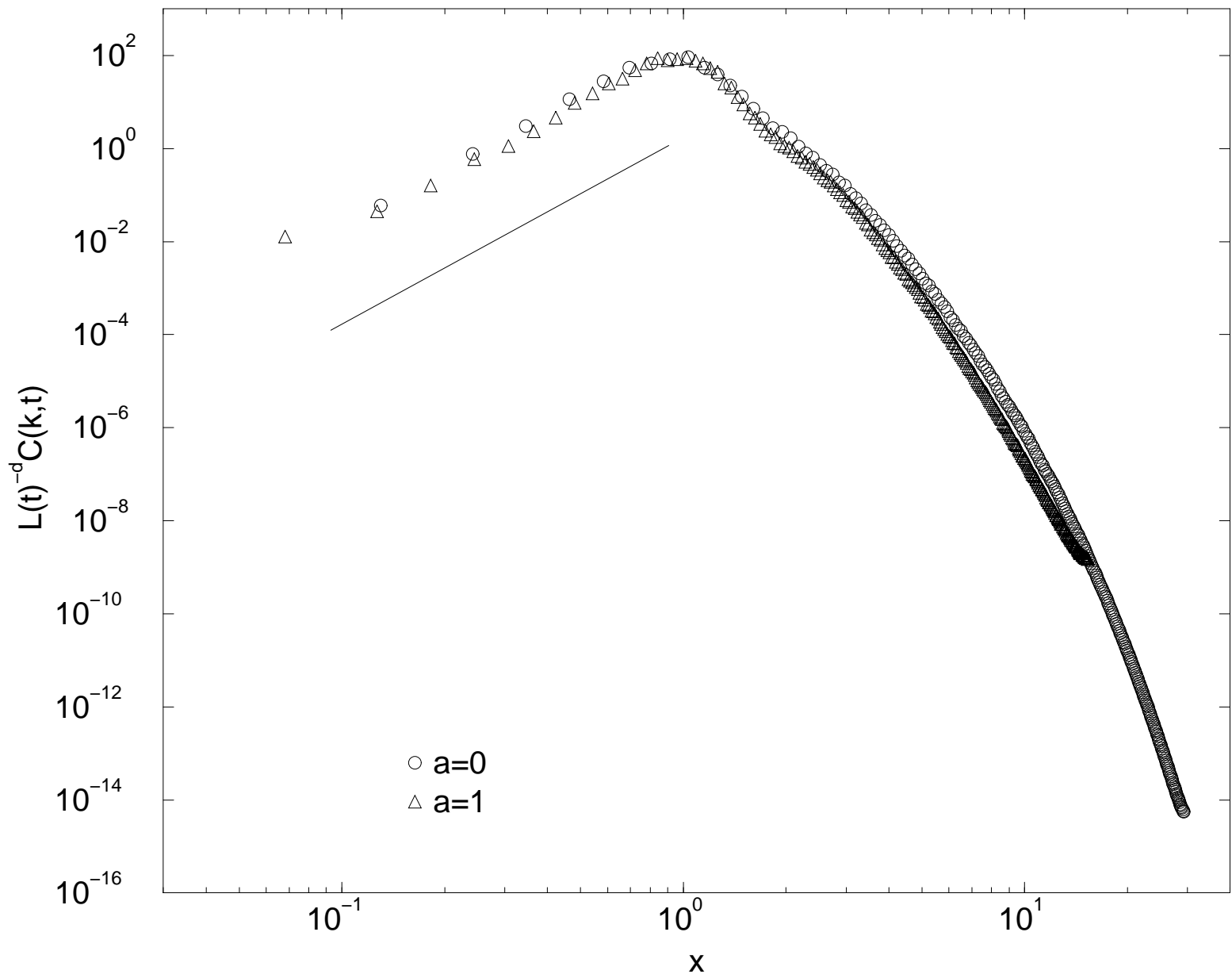
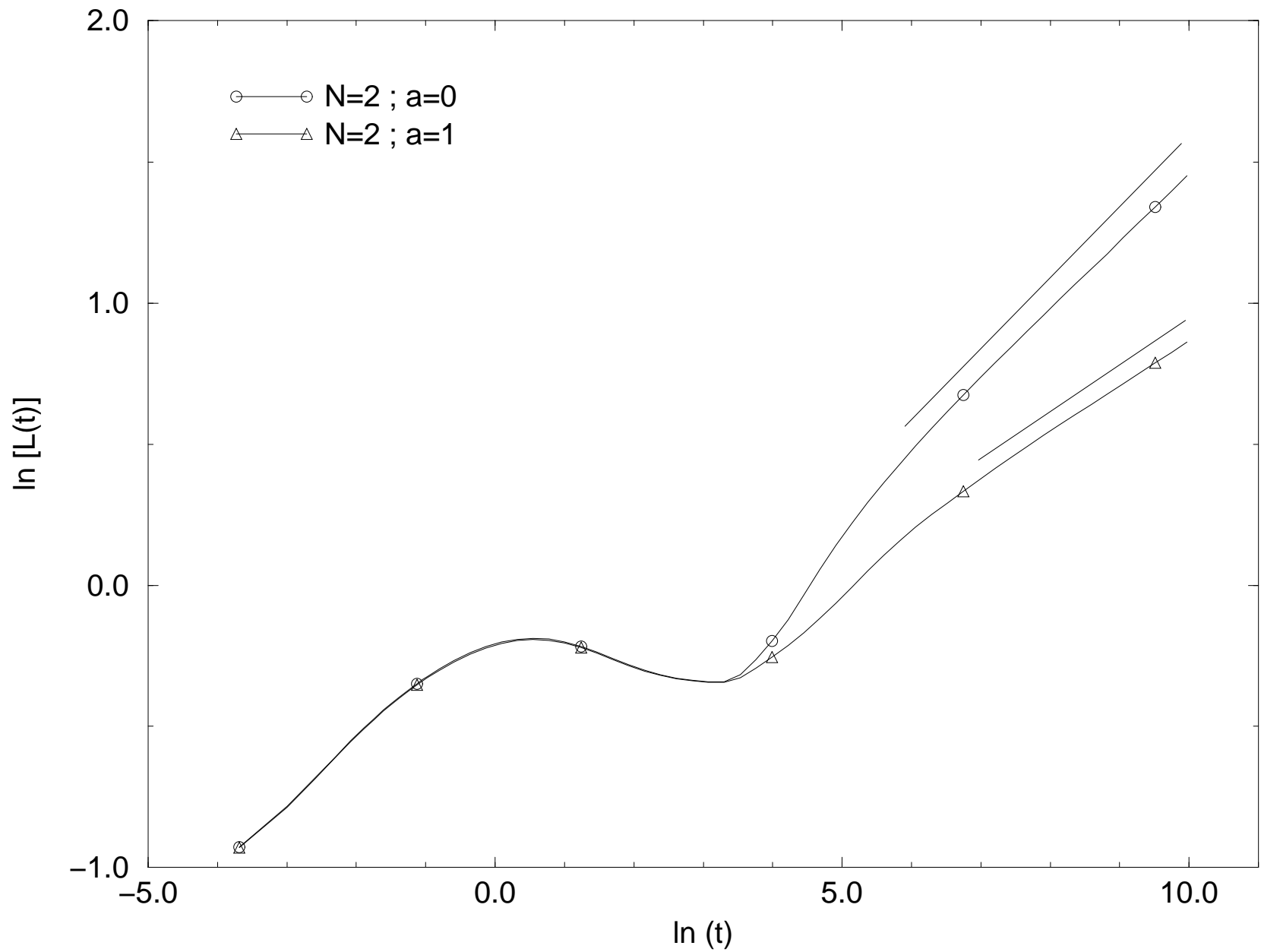
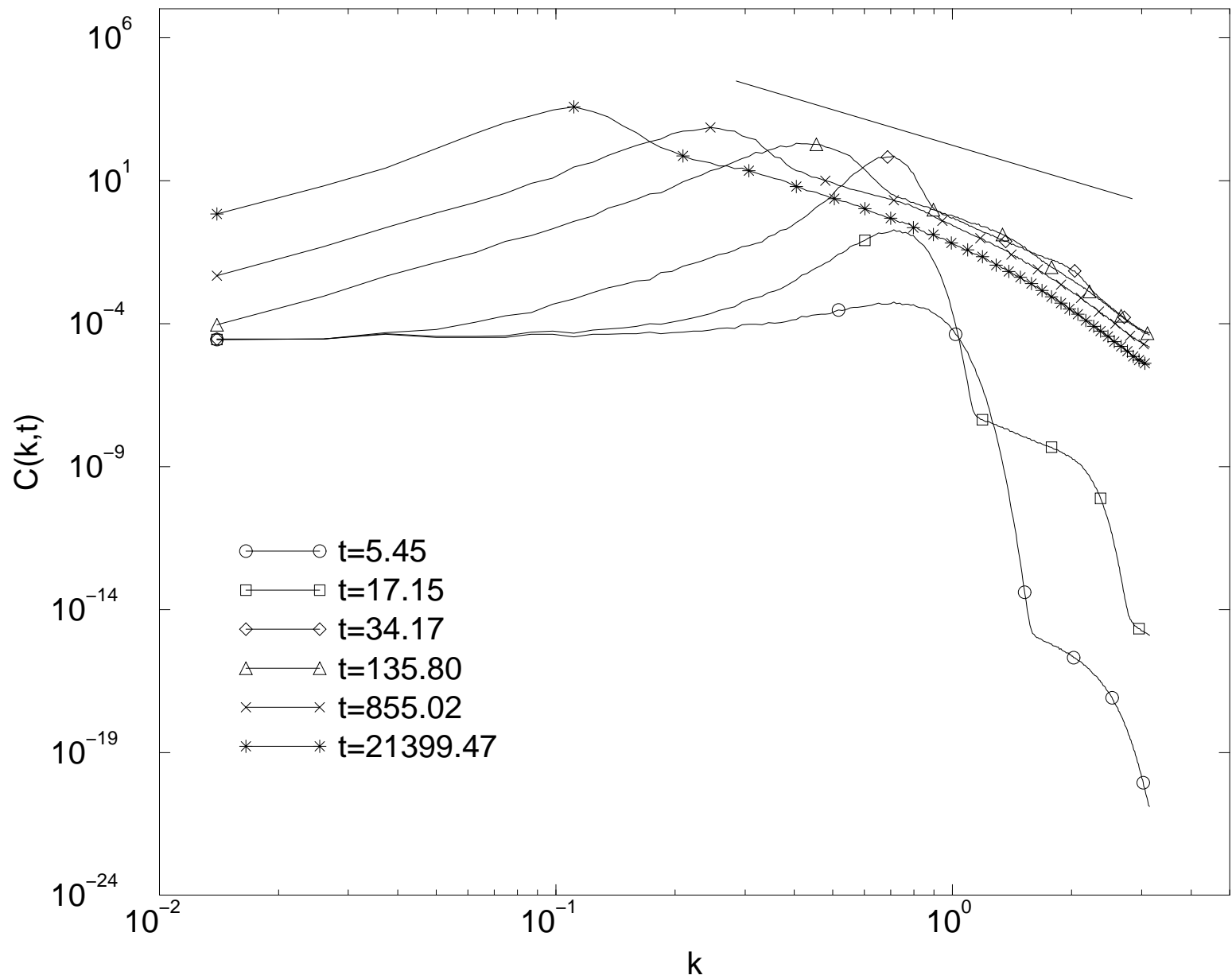
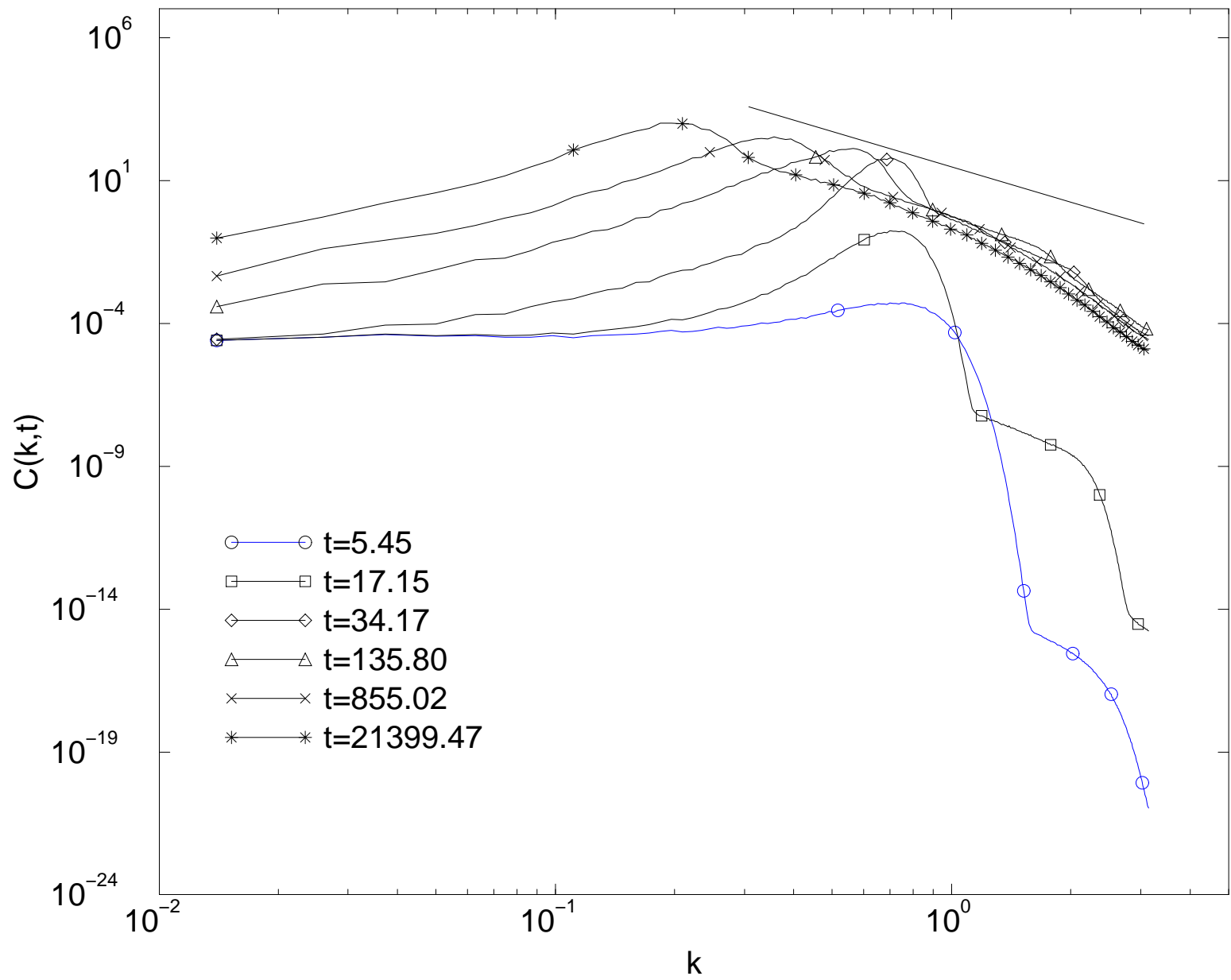
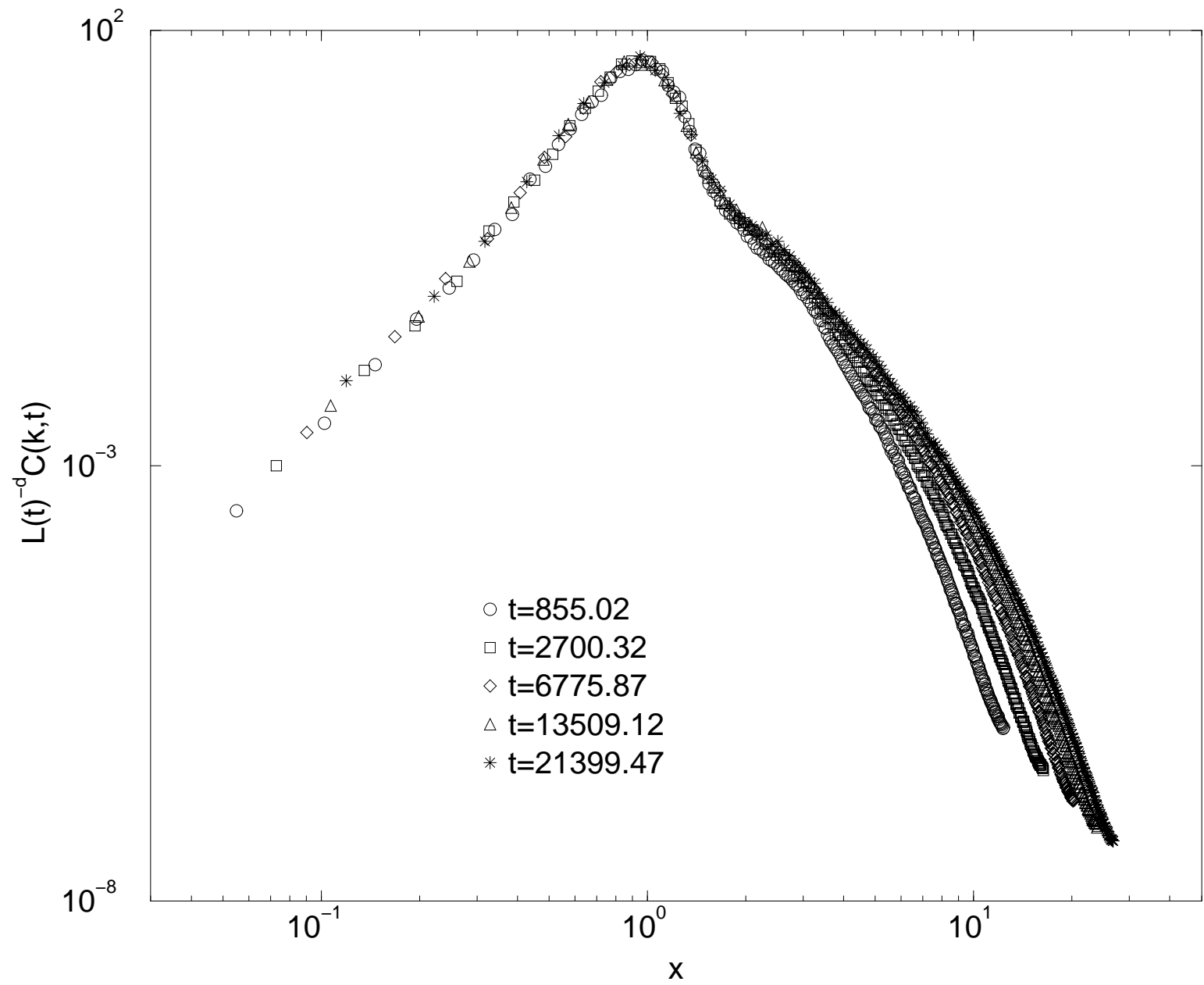
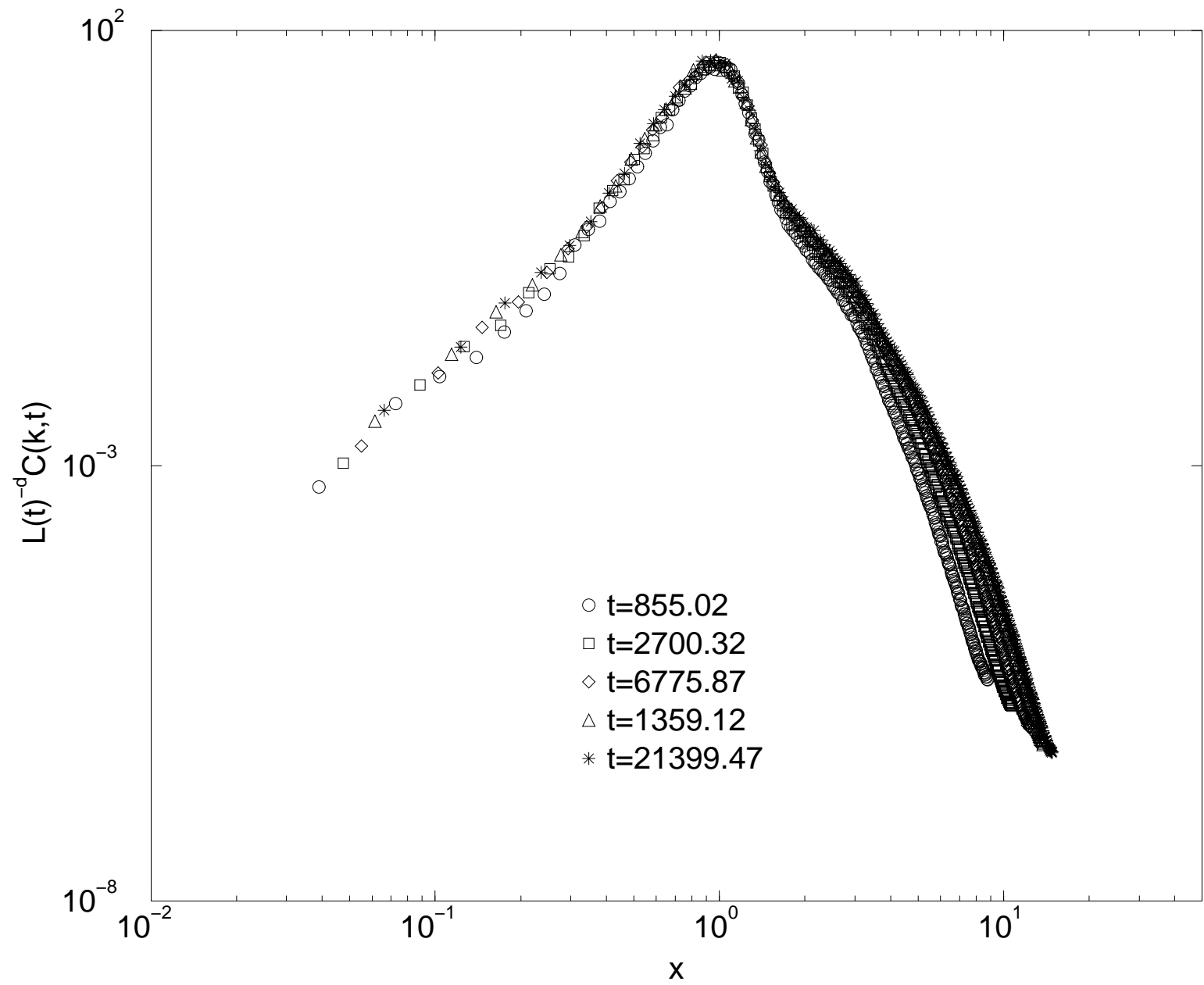
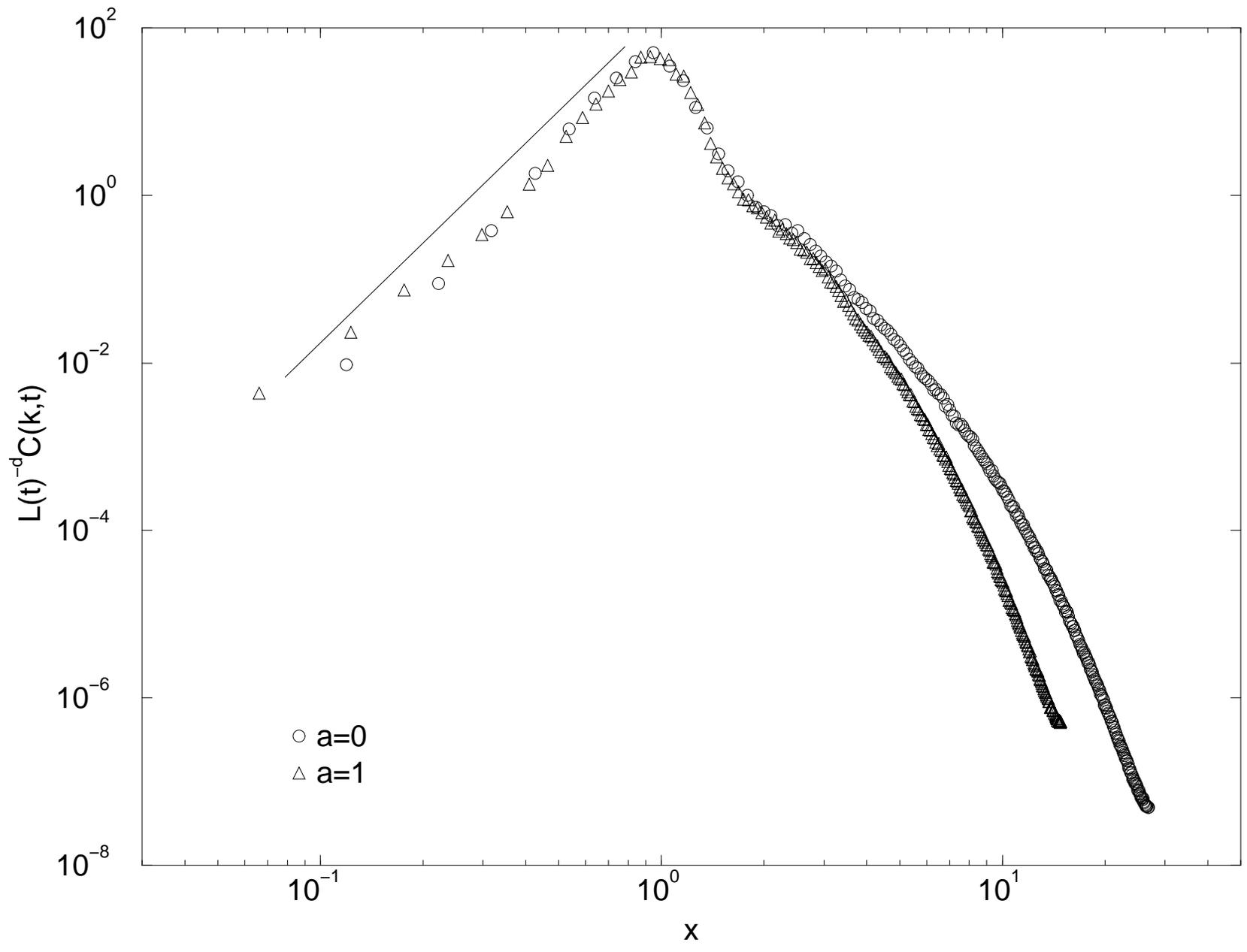
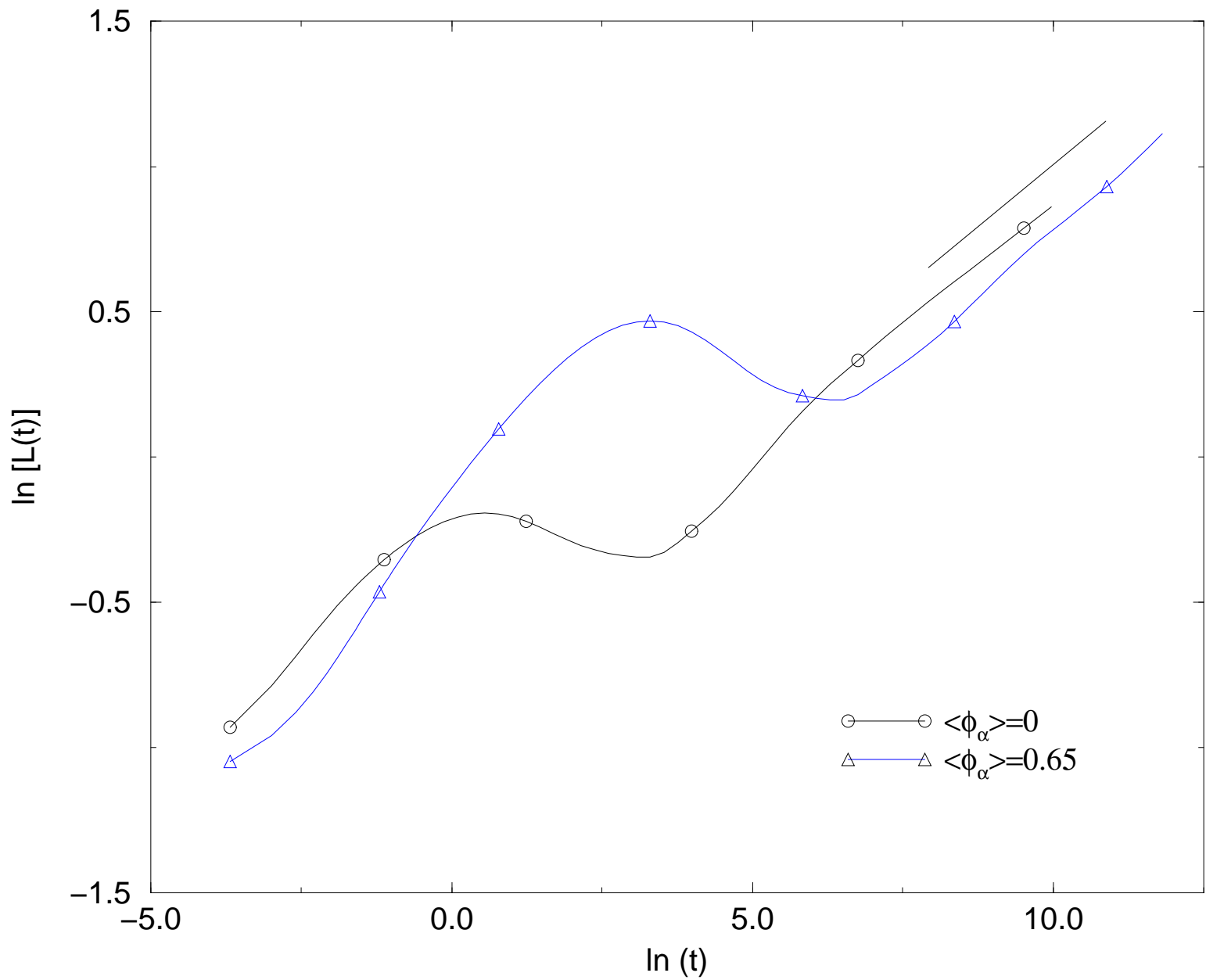
\end{document}